\documentclass[aps,english,notitlepage, superscriptaddress, amsfonts, amssymb, amsmath]{revtex4-1}
\usepackage[T1]{fontenc}
\usepackage[latin9]{inputenc}
\usepackage{amsmath}
\usepackage{amssymb}
\usepackage{graphicx}

\usepackage{babel}
\begin{document}

\title{Broadband telecom transparency of semiconductor-coated metal nanowires:
more transparent than glass }

\author{R. Paniagua-Domínguez }
\affiliation{Instituto de Estructura de la Materia, Consejo Superior de Investigaciones
Científicas, Serrano 121, 28006 Madrid, Spain }

\author{D. R. Abujetas }
\affiliation{Instituto de Estructura de la Materia, Consejo Superior de Investigaciones
Científicas, Serrano 121, 28006 Madrid, Spain }

\author{L. S. Froufe-Pérez }
\affiliation{Instituto de Estructura de la Materia, Consejo Superior de Investigaciones
Científicas, Serrano 121, 28006 Madrid, Spain }

\author{J. J. Sáenz }
\affiliation{Condensed Matter Physics Dept. and Centro de Investigación en Física
de la Materia Condensada (IFIMAC), Universidad Autónoma de Madrid,
Fco. Tomás y Valiente 7, 28049-Madrid, Spain }

\author{J. A. Sánchez-Gil}
\affiliation{Instituto de Estructura de la Materia, Consejo Superior de Investigaciones
Científicas, Serrano 121, 28006 Madrid, Spain }

\email{l.froufe@csic.es}

\begin{abstract}
Metallic nanowires (NW) coated with a high permittivity dielectric
are proposed as means to strongly reduce the light scattering of the
conducting NW, rendering them transparent at infrared wavelengths
of interest in telecommunications. Based on a simple, universal law
derived from electrostatics arguments, we find appropriate parameters
to reduce the scattering efficiency of hybrid metal-dielectric NW
by up to three orders of magnitude as compared with the scattering
efficiency of the homogeneous metallic NW. We show that metal@dielectric
structures are much more robust against fabrication imperfections
than analogous dielectric@metal ones. The bandwidth of the transparent
region entirely covers the near IR telecommunications range. Although
this effect is optimum at normal incidence and for a given polarization,
rigorous theoretical and numerical calculations reveal that transparency
is robust against changes in polarization and angle of incidence,
and also holds for relatively dense periodic or random arrangements.
A wealth of applications based on metal-NWs may benefit from such
invisibility. 
\end{abstract}
\maketitle

\section{Introduction}
In recent years, plasmonic cloaking has received considerable attention
as a mechanism to dramatically reduce the electromagnetic scattering
cross section of an object  \cite{Alu_Adv_Mat_2012,Sanchez_science_2012,Fan2012}.
State of the art technology allows for the fabrication of building
blocks for optical materials which shape, size and composition can
be chosen among a large variety, giving rise to nanostructured engineered
materials with prescribed optical properties. The optical response
of the material in a certain frequency range is often based on the
appearance of resonances in the building blocks; hence, the delicate
interplay among all the parameters controlling the systems response
must be considered. Even relatively simple systems, such as high refractive
index homogeneous dielectric spheres, show interesting light scattering
properties arising from the excitation of single Mie resonances or
superpositions of different ones  \cite{Garcia_Etxarri_2011}. Interestingly,
only in the last years some of those well known properties have been
put forward and measured experimentally \cite{Zhao2009,Greffin2012,Novotny_Nanolett_2013,Luk_yanchuk_Nat_Comm_2013}.

Going one step further, more degrees of freedom can be obtained if
core-shell structures are considered \cite{Norlander_Science_2003}.
The interaction among the modes supported by the different layers
together with size control and materials properties, even presenting
magnetism  \cite{Halas_ACS_Nano_2009}, lead to a certain tunability
of Mie resonances that can result, for instance, in the superposition
of electric and magnetic dipole resonance leading to a certain directivity
control in the light scattering by a submicron particle \cite{Paniagua_2011}.
Suitable configurations of such particles have been proposed for high
directivity, low absorption optical antennas  \cite{Kivshar_ACS_Nano_2012}.
Also, tuning of plasmon resonances in core-shell hybrid metal-dielectric
nanospheres, leads to a large control on the absorption spectrum of
nanoparticles  \cite{Halas_ChemPhys_lett_1998}. Systems presenting
extremely low scattering cross-sections were proposed decades ago
 \cite{Kerker1975} for absorptionless dielectric materials and, more
recently, similar geometries were used as cloaks for metallic spherical
structures  \cite{Chew1976, Engheta_PRE_2005, Engheta_PRE_ERRATUM_2005}. 

Considering available technologies, the use of cylindrical structures
is better suited for many optics and optoelectronics  \cite{Lieber_Mat_today_2006}
applications ranging from nanoantenna emission  \cite{Gomez_Rivas_2012}
to invisible fibers  \cite{Tuniz2010}, cloaking devices  \cite{Alu_NJP_2010,Mundru2012,Jiang_OPEX_2007,Kivshar_OPEX_2013,Smith_OL_2013}, 
superscattering structures \cite{Kivshar_OPEX_2013}, 
or low loss negative index materials  \cite{Sanchez_Gil_2013}. The
most widespread used materials are semiconductors and metals \cite{Nordlander_Acc_Chem_res_2012}.
This kind of structures have been demonstrated to be suitable building
blocks for multifunctional materials. ZnO/Ag nanowire composites  \cite{Kim2013}
have been recently presented as good candidates to fabricate high
electrical conductivity and good optical transparency electrodes with
a potential impact in several industrial applications \cite{Ellmer_Nat_Phot_2012}.
Also growth-controlled semiconductor NWs with different sizes and
geometries have been proposed as means to control light absorption
in photovoltaic applications \cite{Lieber_NL_2012}. 

Different two and three dimensional plasmonic cloaking structures
have been proposed in the literature and experimentally demonstrated
all the way from radio frequency to the optical ranges. Metal coated
dielectric cylinders \cite{Fan2012} have been recently used as a suitable
implementation of a cloaked sensor \cite{Engheta_PRL_2009}. This cloaking
mechanism can present several advantages over other approaches: it
can be achieved with homogeneous and isotropic materials and the underlying
physical mechanism does not depend on resonances, but on the cancellation
of the average polarization density in the object. Hence, fabrication
of actual structures for the infrared and optical ranges should be
eased by the availability of materials and robustness against manufacturing
defects.

Although comprehensive descriptions of the conditions leading to extremely
small light scattering efficiency in hybrid metal-dielectric NWs  \cite{Alu_NJP_2010}
or spheres  \cite{Engheta_PRE_2005, Engheta_PRE_ERRATUM_2005} is
available in the literature, several relevant aspects have been overlooked
so far. In particular, using dielectric coated metallic NWs might
seem similar, if not equivalent, to its inverse structure, namely,
metal coated dielectric NWs. Nevertheless, we shall show that this
is not the case and that the former structure is much more robust
regarding geometrical variations. On the other hand, a deeper analysis
of the near field multiple scattering among transparent structures
is still lacking in the literature. It is the purpose of this manuscript
to perform such an analysis.

In this work, we analyze in detail the conditions required to obtain
small scattering efficiency in a core-shell cylinder for any metal
or dielectric combination throughout remarkably broad bands in the
IR spectral region relevant to telecommunications \cite{Paschotta_2008}. By the use of a
simple model based on the quasi-static approximation with radiative
corrections to the polarizability of a core-shell cylinder \cite{Albaladejo2010},
we obtain general properties required to achieve transparency in realistic
structures  \cite{Alu_NJP_2010}. We also check our predictions against
a more accurate model based on Mie theory for coated cylinders \cite{Bohren_Huffman,Shah1970}.
We find that, under rather general conditions, metal nanowires with
high refractive index coatings can show a transparency region which
is more robust against fabrication defects (size polydispersity) than
metal coated fibers. Also, it is shown that it is possible to obtain
up to three orders of magnitude lower scattering efficiency, compared
with raw metal cylinders, in a band as wide as 20\% of the central
frequency, and with realistic materials (Si coated Ag wires) in the
infrared. The transparency condition is also quite robust regarding
the angle of incidence and polarization of the incoming signal. It
is shown that the near field scattering is extremely weak in the transparency
region. Hence, the coupling through evanescent modes among cylinders
is essentially negligible. As a consequence, a high density assembly
of appropriately designed NWs present an extremely low scattering
efficiency. For the sake of comparison,we show that a properly designed
coating for a metal NW with a given diameter, can present smaller
scattering efficiency than a glass slab of equivalent thickness.

\section{Transparency conditions}
For normal incidence and transverse
magnetic (TM) polarized plane waves, the electric field in the electrostatic
approximation is constant and parallel to the cylinder axis. The quasi-static
polarizability in this case is
\begin{equation}
\alpha_{0}^{(TM)}=A\left[\left(\epsilon_{c}-\epsilon_{h}\right)R^{2}+\left(\epsilon_{s}-\epsilon_{h}\right)\left(1-R^{2}\right)\right]\label{eq:2_10}
\end{equation}
where $A=\pi R_{s}^{2}$ is the total cross section of the cylinder
and $R\equiv R_{c}/R_{s}$ is the ratio of the core to shell radii.
Throughout this paper, sub-index $h$ refers to the host medium, $s$
to the shell medium, and $c$ to the core medium; to depict the structures, we 
shall use core-material@shell-material notation (for instance Ag@Si denotes a silver core NW coated with silicon). The quasi-static
polarizability given by Eq. (\ref{eq:2_10}) can be interpreted
as the cross-section averaged susceptibility. Interestingly, the quasi-static
polarizability (\ref{eq:2_10}) is invariant if we exchange $\epsilon_{c}$
with $\epsilon_{s}$, and $R^{2}$ with $\left(1-R^{2}\right)$. In
fact, the quasi-static polarizability is proportional to the cross-section
average of the polarizability density in the quasi-static approximation.
Hence if the relative amount of metal and dielectric is kept constant,
any spatial distribution of both components would lead to the same
polarizability. This fact establishes a correspondence between metal@dielectric
and dielectric@metal NWs.  

A radiative correction to the quasi-static polarizability must be
included in order to obtain a correct energy balance between scattering,
absorption and extinction in the scattering process. The dynamic polarizability
in the small particle approach is $\alpha^{(TM)}=\alpha_{0}^{(TM)}/\left(1-ik_{h}^{2}\alpha_{0}^{(TM)}/4\right)$
from which the scattering efficiency can be obtained as
\begin{equation}
Q_{scat}^{(TM)}=\left|\alpha^{(TM)}\right|^{2}k_{h}^{3}/\left(8\epsilon_{h}^{2}R_{s}\right).\label{eq:2_15}
\end{equation}

From Eq. (\ref{eq:2_10}), the scattering efficiency presents
a minimum for a size ratio $R=R_{tr}^{(TM)}$ such that
\begin{equation}
\left(R_{tr}^{(TM)}\right)^{2}=\frac{\epsilon_{h}-\epsilon'_{s}}{\epsilon'_{c}-\epsilon'_{s}}\label{eq:2_20}
\end{equation}
where primes in the above expression indicate the real part of the
permittivity (the host material is assumed to be absorptionless).
When this transparency condition is fulfilled, the averaged polarization
density in the cross section of the cylinder vanishes.

Several cases can be discussed depending on the nature of the materials.
For metal@dielectric cylinders, we consider $\epsilon'_{c}<0$ and
$\epsilon'_{s}>\epsilon_{h}>0$, leading to $R_{tr}^{TM}<1$; i.e.
any metal core and dielectric shell with refractive index larger than
the host medium presents an optimum size ratio $R_{tr}^{TM}$ for
transparency. In the inverse case, dielectric@metal, there is an optimum
size ratio $R_{tr}$ if the dielectric core presents a refractive
index higher than the host medium. On the other hand, pure dielectric
structures can also show transparency if $\epsilon'_{s}>\epsilon_{h}>\epsilon'_{c}$
or $\epsilon'_{c}>\epsilon_{h}>\epsilon'_{s}$. Obviously, these latter
two conditions can not be fulfilled in vacuum. For purely metallic
structures, there is no optimum ratio even in the absorptionless limit.

Analogously, we analyze also the TE polarization. In this case, considering
the matching of the electric field at the boundaries, we obtain a
quasi-static polarizability
\begin{equation}
\alpha_{0}^{(TE)}=2A\frac{\left(\epsilon_{s}-\epsilon_{h}\right)\left(\epsilon_{s}+\epsilon_{c}\right)+\left(\epsilon_{c}-\epsilon_{s}\right)\left(\epsilon_{h}+\epsilon_{s}\right)R^{2}}{\left(\epsilon_{s}+\epsilon_{h}\right)\left(\epsilon_{s}+\epsilon_{c}\right)-\left(\epsilon_{c}-\epsilon_{s}\right)\left(\epsilon_{h}-\epsilon_{s}\right)R^{2}}.\label{eq:2_30}
\end{equation}
The polarizability with radiative correction reads in this case, $\alpha^{(TE)}=\alpha_{0}^{(TE)}/\left(1-ik_{h}^{2}\alpha_{0}^{(TE)}/8\right)$,
and the scattering efficiency, 
\begin{equation}
Q_{scat}^{(TE)}=\left|\alpha^{(TE)}\right|^{2}k_{h}^{3}/\left(16\epsilon_{h}^{2}R_{s}\right)\textrm{.}\label{eq:2_35}
\end{equation}
The optimum size ratio $R=R_{tr}^{(TE)}$ that minimizes the scattering
is reached for vanishing quasi-static polarizability; in this case
the optimum size is
\begin{equation}
\left(R_{tr}^{(TE)}\right)^{2}=\frac{\left(\epsilon_{h}-\epsilon'_{s}\right)\left(\epsilon_{c}'+\epsilon_{s}'\right)}{\left(\epsilon{}_{h}+\epsilon'_{s}\right)\left(\epsilon_{c}'-\epsilon_{s}'\right)}=\left(R_{tr}^{(TM)}\right)^{2}\frac{\left(\epsilon_{s}'+\epsilon_{c}'\right)}{\left(\epsilon_{s}'+\epsilon_{h}\right)}.\label{eq:2_40}
\end{equation}

Both Eqs. (\ref{eq:2_20}) and (\ref{eq:2_40}) are in full agreement
with the ones obtained in  \cite{Alu_NJP_2010}. Contrary to the
TM polarization case, Eq. (\ref{eq:2_40}) for TE polarization
is more restrictive regarding materials. For metal@dielectric it is
not always possible to obtain an optimum size ratio minimizing light
scattering in TE polarization. If $\left|\epsilon'_{c}\right|>\epsilon'_{s}$
, which is a typical case in the optical and infrared for many metals
and available dielectrics, the ratio given by Eq. (\ref{eq:2_40})
is negative and hence the solution is unphysical. 

On the other hand, it is clear from Eqs. (\ref{eq:2_20}) and (\ref{eq:2_40})
that it is not possible to minimize the scattering efficiency for
a single core-shell structure at both TM and TE polarizations at the
same wavelength. However, we will show later on that the TM transparency
condition spectrally overlaps with a region of weak TE scattering
efficiency, thus leading to highly polarization-independent transparency.

Nevertheless, the situation changes for dielectric@metal structures.
An optimum size ratio $R_{tr}^{(TE)}$ can be reached with realistic
materials. In this case, we can also expect the appearance of a localized
surface plasmon (LSP) resonance near the condition of minimum scattering
efficiency. It can be shown that, in the limit $\epsilon'_{c}\gg\epsilon_{h}$
(for instance, semiconductor core) and $\left|\epsilon'_{s}\right|^{2}\gg\epsilon'_{c}\epsilon_{h}$
(common with noble metals as shell material), the wavelength of the
LSP resonance $\lambda_{lsp}$ is shifted with respect to the chosen
wavelength for transparency $\lambda_{tr}$ by an amount $\left.\left(d\epsilon'_{s}/d\lambda\right)\right|_{\lambda=\lambda_{tr}}\left(\lambda_{lsp}-\lambda_{tr}\right)\simeq\epsilon_{h}/\epsilon_{c}$.
Hence, metal shells with large chromatic dispersion and high refractive
index dielectric cores can lead to a spectrally sharp LSP-to-transparency
transition. We postpone the study of its implications to a further
work.

For dielectric@metal structures at fixed wavelength, there is also
an optimum size ratio $R=R_{lsp}^{(TE)}$ for which the LSP resonance
matches the selected wavelength. Again considering the divergence
of the quasi-static polarizability we obtain 
\begin{equation}
\Delta R\equiv\frac{R_{tr}^{(TE)}-R_{lsp}^{(TE)}}{R_{tr}^{(TE)}+R_{lsp}^{(TE)}}=\left|\frac{\epsilon_{h}}{\epsilon_{s}'}\right|\label{eq:2_50-1}
\end{equation}
For instance, at $\lambda=1550\textrm{ nm}$, a Si@Ag cylinder in
vacuum presents a ratio $\Delta R=1/130\simeq7.7\times10^{-3}$. Hence,
relatively small deviations from the optimum size ratio would have
a large impact in the scattering efficiency spectrum. 

\begin{figure}[htbp]
\centering\includegraphics[width=1\textwidth]{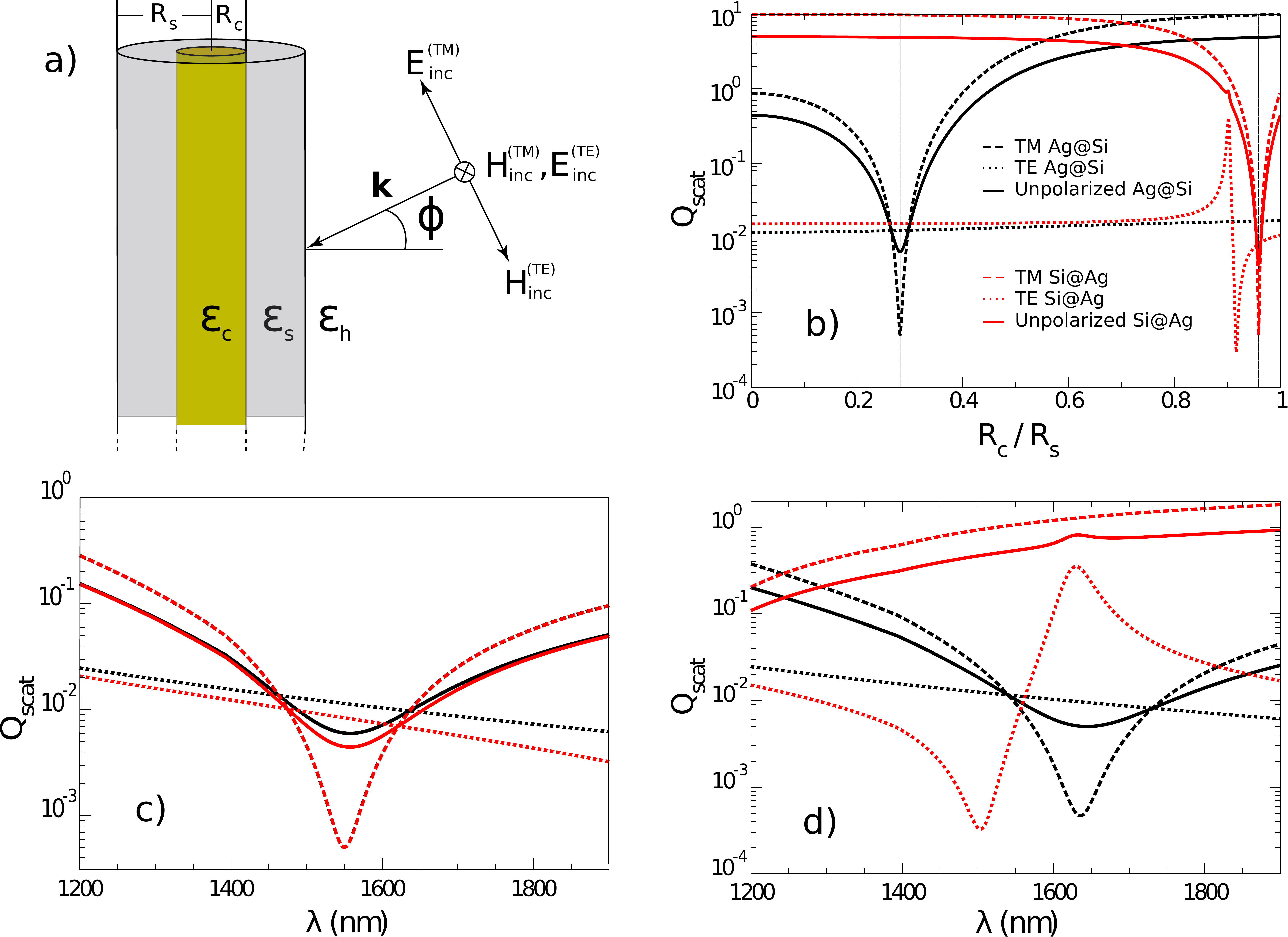}							
\caption{\label{fig:1}(a) Sketch of the system under consideration
together with the relevant parameters. (b-d), scattering efficiency
$Q_{scat}$ as obtained from the quasi-static approximation. TM polarization
(dashed line), TE polarization (dotted line) and unpolarized (solid
line) radiation is considered for both Ag@Si (black curves) and Si@Ag
(red curves) structures. (b) $Q_{scat}$ at a constant wavelength
$\lambda=1550\textrm{ nm}$ is plotted as a function of the core to
shell radii ratio. (c), the spectra for the different polarizations
are presented. The size ratio $R_{c}/R_{s}$ is fixed in such a way
that $Q_{scat}$ is minimized in TM polarization for each structure.
(d) $Q_{scat}$ spectra are presented when the core radius is reduced
by 5\%. }
\end{figure}

To illustrate the results of this section, we plot in Fig. \ref{fig:1} 
the scattering efficiency from Eqs. (\ref{eq:2_15}) and (\ref{eq:2_35})
of core-shell nano-cylinders considering TM and TE polarizations and
unpolarized radiation for both metal@dielectric and dielectric@metal
configurations [sketched in Fig. \ref{fig:1}(a)]. We have chosen silicon
as a suitable semiconductor in the infrared, and silver as the metallic
component. In the spectral region of interest, we use $\epsilon_{Si}=12.25$,
whereas $\epsilon_{Ag}$ is taken from  \cite{Johnson_Christy_1972}.

In Fig. \ref{fig:1}(b), $Q_{scat}$ is shown as a function of the size ratio
$R=R_{c}/R_{s}$ at a working wavelength $\lambda=1550\textrm{ nm}$
($\epsilon_{Ag}\simeq-130+3.3i$). As predicted by Eq. (\ref{eq:2_20}),
$Q_{scat}$ for TM polarization (dashed lines) presents a sharp minimum
when the cross-sectional average of the electric polarization density
vanishes. In the case under study, this condition is fulfilled when
the Ag to Si volume ratio is $V_{Ag}/V_{Si}=\left(\epsilon'_{Si}-1\right)/\left(\epsilon'_{Ag}-1\right)\simeq0.086$,
which corresponds to $R_{tr}^{TM}\simeq0.28$ for a Ag@Si NW, and
$R_{tr}^{TM}\simeq0.96$ for a Si@Ag NW. Both values of the optimum
size ratio for transparency are represented by vertical lines in Fig. \ref{fig:1}(b).

For TE polarization, as predicted by Eq. (\ref{eq:2_40}), there is
no optimum size ratio minimizing $Q_{scat}$ at the working wavelength
for Ag@Si NWs, but it shows a weak monotonic increase with $R$.
On the contrary, for Si@Ag NWs we predict a minimum in the $Q_{scat}$
at $R_{tr}^{TE}\simeq0.917$. Nevertheless, at a slightly smaller
value of the size ratio, as predicted by Eq. (\ref{eq:2_50-1}), a
LSP resonance occurs. At $\lambda=1550\textrm{ nm}$, the size ratio
at plasmon resonance given by Eq. (\ref{eq:2_50-1}) is $R_{plsp}^{TE}=R_{tr}^{TE}129/131$,
which is of the order of 1.5\% smaller than the optimum size ratio
for transparency.

The scattering efficiency spectrum is represented in the region of
interest for core-shell dimensions satisfying the optimum condition
[see Fig. \ref{fig:1}(c)] for TM polarization. As expected, the scattering
spectra exhibit a pronounced minimum at the working wavelength of
$\lambda=1550\textrm{ nm}$ for this polarization. In fact, in this
case, $Q_{scat}$ spectra for both Ag@Si and Si@Ag NWs are exactly
the same. Moreover, as can be seen in Fig. \ref{fig:1}(c), the spectra for
Ag@Si and Si@Ag for unpolarized radiation are very similar, with only
a small difference due to the TE polarization component.

We might conclude that, Ag@Si NWs behave essentially as Si@Ag if the
metal/dielectric volume fraction is kept to the optimum value for
achieving transparency at the selected wavelength. Nevertheless, as
indicated above, small changes in the size ratio of the structure
might lead to large variations in the $Q_{scat}$ spectrum in TE polarization
for dielectric@metal structures. In deed, as can be observed in Fig. \ref{fig:1}(d),
this is the case when the size ratio is diminished by 5\%. The spectrum
corresponding to Ag@Si red shifts by about 5.6\%, while the spectrum
corresponding to Si@Ag varies dramatically, not showing a minimum
and increasing its value by a large amount in the whole considered
band. This is due to the fact that, at size ratios close to the optimum
one, the spectrum of Si@Ag nanowires shows a LSP resonance. In this
regard, transparency in metal@dielectric structures is expected to
be more robust against structural variations.

From the quasi-static polarizability approximations, we can also obtain
useful information regarding the minimum value of the scattering efficiency
and bandwidth. We focus on metal@dielectric structures in TM polarization
since this structure is, as stated above, more robust against fabrication
errors, and the TM polarization controls the scattering spectrum while
the TE polarization adds a slowly varying background. The transparency
condition given by Eq. (\ref{eq:2_20}) links the size ratio of the optimized structure
with the real parts of the polarizabilities of the different materials.
The chromatic dispersion and the absorption controls the bandwidth
of the transparency region and the minimum value of the scattering
efficiency respectively.

If we compare the scattering cross-length in TM polarization of the
optimized Ag@Si structure with the bare Ag core, we obtain, after
some algebra, a scattering cross-length ratio (in this case c=Ag,
s=Si and h=vacuum)
\begin{equation}
\frac{\sigma_{s}^{(TM,Ag@Si)}}{\sigma_{s}^{(TM,Ag)}}\simeq\left|1+\frac{\epsilon_{s}-\epsilon_{h}}{\epsilon_{c}-\epsilon_{h}}\frac{\epsilon_{c}^{(tr)}-\epsilon_{h}^{(tr)}}{\epsilon_{h}^{(tr)}-\epsilon_{s}^{(tr)}}\right|^{2}\label{eq:2_60}
\end{equation}
where $(tr)$ denotes the polarizability value taken at the transparency
condition. The optimized structure presents less scattering than the
bare one if this ratio is smaller than unity. If we further consider
that both host and shell do not show chromatic dispersion, and that
the permittivity of the metal is much larger than the host medium
one, $\left|\epsilon_{c}\right|\gg\epsilon_{h}$, this ratio simplifies
to
\begin{equation}
\frac{\sigma_{s}^{(TM,Ag@Si)}}{\sigma_{s}^{(TM,Ag)}}\simeq\left|1-\frac{\epsilon_{c}^{(tr)}}{\epsilon_{c}}\right|^{2}\label{eq:2_70}
\end{equation}
hence, considering relatively small losses in the metal, the core-shell
structure scatters less efficiently in a spectral window where $\left|\epsilon_{c}\left(\lambda\right)\right|>2\left|\epsilon_{c}\left(\lambda_{tr}\right)\right|$.
In the case of Ag@Si NW in vacuum, this transparency region starts
at a wavelength $\lambda\simeq1100\textrm{ nm}$ for an optimized
structure with a scattering minimum at $\lambda_{tr}=1550\textrm{ nm}$.
At longer wavelengths, the bare silver structure scatters light more
efficiently than the core-shell one.

At the minimum scattering wavelength, the scattering efficiency in
this configuration can be estimated to be
\begin{equation}
Q_{scat}^{(TM)}\left(\lambda=\lambda_{tr}\right)\simeq\frac{\pi^{5}}{\sqrt{\epsilon_{h}}}\left(\frac{R_{s}}{\lambda_{tr}}\right)^{3}\left(\epsilon''_{c}\frac{\epsilon_{h}-\epsilon'_{s}}{\epsilon'_{s}-\epsilon'_{c}}\right)^{2},\label{eq:2_80}
\end{equation}
where $\epsilon'=\textrm{Re}\left(\epsilon\right)$ and $\epsilon''=\textrm{Im}\left(\epsilon\right)$.
In the example shown in Fig. 1(b), we have $Q_{scat}^{(TM)}\left(\lambda=1550\textrm{ nm}\right)\simeq5.10\times10^{-4}$,
which is of the order of 1\% larger than the numerically obtained
value.

From Eq. (\ref{eq:2_80}) it is clear that the scattering efficiency
at the transparency wavelength is determined by the absorption level
in the metal (imaginary part of the permittivity) and is smaller the
smaller the structure is.

\section{Single Ag@Si NW transparency: Extended Mie calculations}
To check the predictions obtained in the previous section, which were
derived by using a quasi-static approximation, we shall focus in this
section on Ag@Si cylinders, rigorously calculating scattering cross
sections by Mie expansion. We consider a multilayered concentric cylindrical
structure; the fields at each layer are suitably expanded as a sum
of cylindrical vector harmonics. Solving for the matching conditions
at each interface and expanding the incoming plane wave in the same
basis, provides an analytic solution of the scattering problem for
any incoming radiation direction and polarization  \cite{Kerker1961,Shah1970}. 

\begin{figure}[htbp]
\centering\includegraphics[width=1\textwidth]{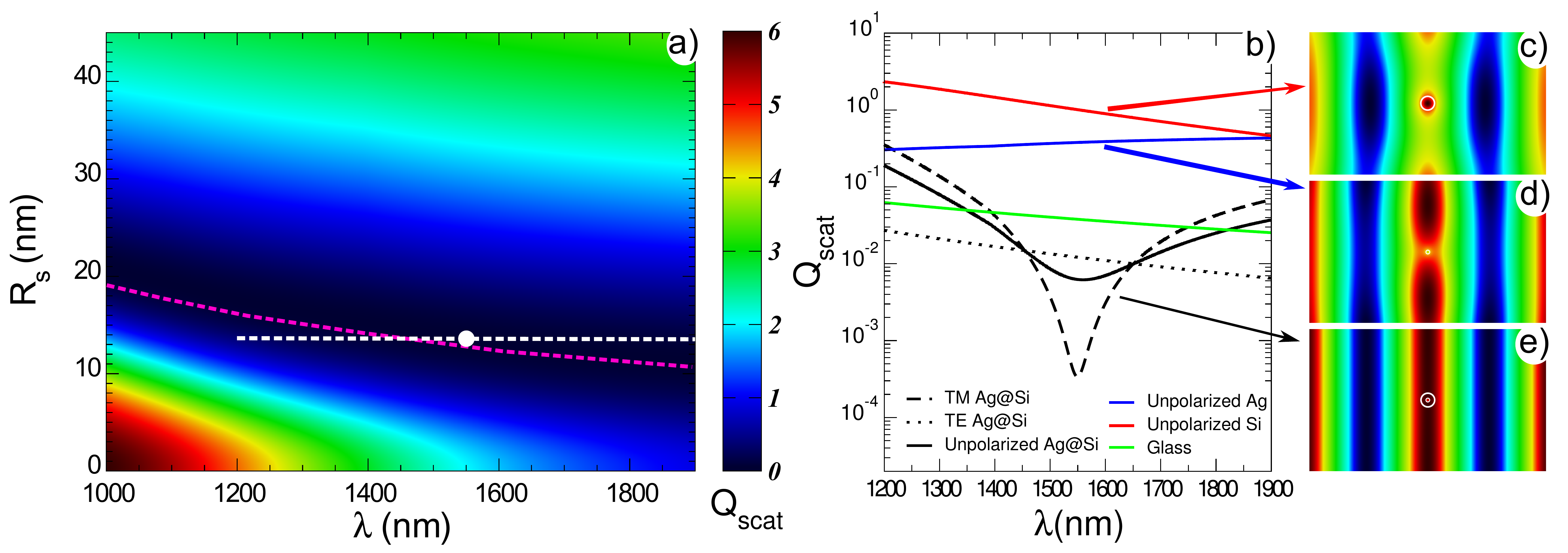}
\caption{\label{fig:2}(a) Scattering efficiency spectra as
a function of the core (silver) radius for a fixed shell (silicon)
outer radius $R_{s}=45\textrm{ nm}$. (b) Scattering efficiency spectra
for a Ag@Si core-shell NW ($R_{s}=13.6\textrm{ nm}$, $R_{s}=13.6\textrm{ nm}$)
and different polarizations: TM (black dashed), TE (black dotted)
and unpolarized (black solid line). For the sake of comparison we
also represent $Q_{scat}$ for a homogeneous Ag NW (red curve, $R=13.6\textrm{ nm})$,
a homogeneous Si NW (blue curve, $R=45\textrm{ nm})$, and for a homogeneous
90 nm thick glass slab (refractive index n=1.45). (c-e) Maps of the
electric field along the cylinder axis direction in TM polarization
at a working wavelength of $\lambda=1550\textrm{ nm}$ (corresponding
to the minimum of the black curve in b). (c) Bare silver NW. (d) Homogeneous
silicon NW. (e) Ag@Si NW.}
\end{figure}

We are interested in the telecommunications range and hence we take
a band centered at $\lambda=1550\textrm{ nm}$. In Fig. \ref{fig:2}(a),
a color map of the scattering efficiency ($Q_{scat}$) of a Ag@Si
NW is shown. The radius of the shell is kept constant at $R_{s}=45\textrm{ nm}$,
while the core radius varies from $R_{c}=0\textrm{ nm}$ (pure Si
NW) to $R_{c}=45\textrm{ nm}$ (pure silver NW). The vacuum wavelength
varies in the range $1000\textrm{ nm}\leq\lambda\leq1900\textrm{ nm}$.
For a core radius $R_{s}\simeq14\,\textrm{nm}$ (white dashed line),
the NW presents a scattering efficiency spectrum much weaker than
those of either the homogeneous Si or Ag NWs alone [Fig. \ref{fig:2}(b)]
by one or two orders of magnitude, in the range $1400\textrm{ nm}\leq\lambda\leq1900\textrm{ nm}$,
and for unpolarized radiation. The exact value value of the optimum $R_{s}$ for this system is only
7.4\% larger than the quasi-static prediction of Eq. (\ref{eq:2_20}),
which supports the validity of the approximate transparency
condition for subwavelength NW dimensions. In fact, for the chosen NW diameter, materials and
wavelengths, contributions from higher order multipoles than the ones in the quasi-static approximation
are essentially negligible \cite{Alu_NJP_2010}. For the TM polarization, estudied in detail in this section, only the monopole mode
$n=0$ \cite{Alu_NJP_2010,Kivshar_OPEX_2013} significantly contributes to the scattering cross section, while for the TE mode, only the
dipolar mode ($n=\pm 1$) contributes to scattering.

In order to stablish a comparison with a simple lossless scattering system, we calculate the
scattering efficiency of a thin glass slab (refractive index $n\simeq1.45$). In this case, $Q_scat$
can be obtained through the Fresnel coefficients of the structure. If $t$ is the transmission
coefficient as a function of the wavelength, $Q_scat=2\left( 1-\textrm{Re}(t)\right)$. Although basic 
antireflecting coatings might be used to minimize the scattering of the glass slab, it is
remarkable that the optimized Ag@Si NW scatters less efficiently than a 90 nm thick glass slab despite 
the metallic character of the core.

In Figs. \ref{fig:2}(c)--\ref{fig:2}(e), a color map of the electric field along the direction
parallel to the cylinder axis is shown for three different cases at
$\lambda=1550\textrm{ nm}$. The incoming signal is a TM polarized
plane wave at normal incidence. The electric field is represented
in a plane perpendicular to the cylinder. In Fig. \ref{fig:2}(c), a single
homogeneous silicon NW is considered (sketched as a circle at the
center of the Fig.) of radius $R=45\textrm{ nm}$. As can be seen
in the Fig., the wave fronts are largely altered due to the interference
of the incoming wave with the scattered field. A similar behavior
is obtained when a homogeneous silver NW of radius $R=13.6\textrm{ nm}$
is considered [Fig. \ref{fig:2}(d)]. Nevertheless, as shown in Fig. \ref{fig:2}(e),
when the same silver NW is covered by a $31.4$ nm thick silicon shell,
the total electric field does not appreciably differ from the plane
wave even in the close proximity of the NW surface. Hence the scattering
efficiency of the core-shell NW is strongly reduced in this situation.

\begin{figure}[htbp]
\centering\includegraphics[width=1\textwidth]{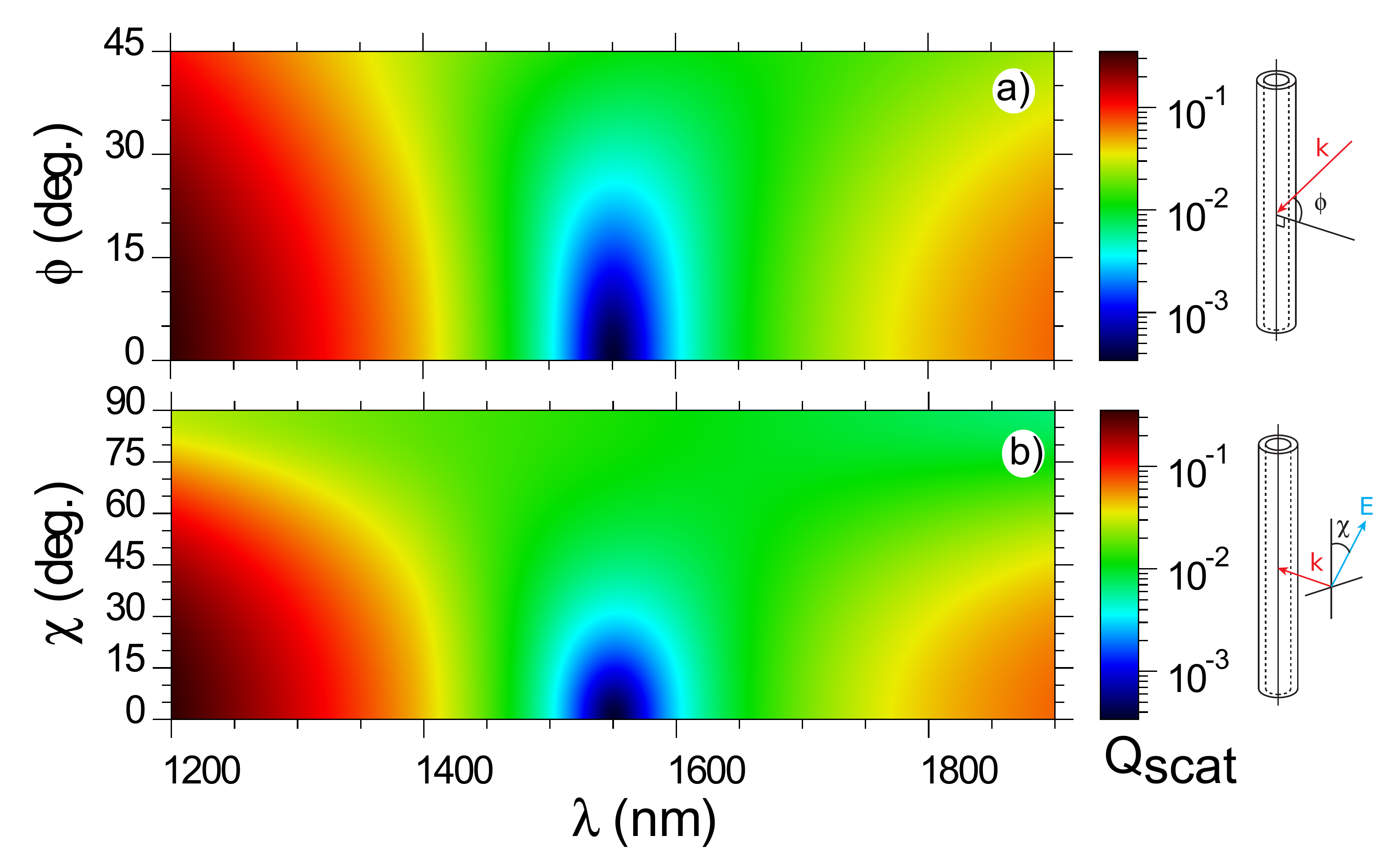}
\caption{\label{fig:3}(a) Scattering efficiency spectra for
a Ag@Si core-shell NW ($R_{c}=13.6\textrm{ nm}$ and $R_{s}=45\textrm{ nm}$)
as a function of (a) the angle of incidence ($\phi$) for TE polarized
light and (b) the polarization angle ($\chi$) at normal incidence
for the same structure.}
\end{figure}

In order to assess the robustness of transparency regarding the incoming
signal, we now consider different incidence angles and polarizations.
In Fig. \ref{fig:3}(a), an incoming TM polarized plane wave impinges on
the core-shell cylinder. In this case we keep the geometry of the
system constant ($R_{s}=45\textrm{ nm}$, $R_{s}=13.6\textrm{ nm}$).
The scattering efficiency keeps at very low values, of the order of
$10^{-2}$ in a band centered at $\lambda=1550\textrm{ nm}$, with
band-width is of the order of $200$ nm for angles deviating from
normal incidence as much as 45 degrees. Hence, we can conclude that
the transparency band is very robust against variations in the angle
of incidence.

Regarding polarization, it is rather clear that the optimum polarization
corresponds to TM. Nevertheless, as shown in Fig. \ref{fig:3}(b), low scattering
efficiency regions, for normal incidence and linearly polarized light,
are quite robust against variations in the polarization state of the
incoming light. In this Fig. we show a color map of the scattering
efficiency as a function of the angle $\chi$ between the incoming
electric field and the cylinder axis. If $Q_{\parallel}$ and $Q_{\perp}$
are the scattering efficiencies for TM and TE polarizations respectively,
the scattering efficiency for a general linearly polarized plane wave
at normal incidence is $Q_{scat}=Q_{\parallel}\cos^{2}\left(\chi\right)+Q_{\perp}\sin^{2}\left(\chi\right)$.
Again, a band between $\lambda=1400\textrm{ nm}$ and $\lambda=1850\textrm{ nm}$
shows a scattering efficiency $Q_{scat}\le0.05$ for any polarization
state.

\section{Transparency of NW assemblies}
We now address the effects
due to the multiple scattering of light in arrangements of NWs relatively
close to each other. In principle, multiple light scattering can lead
to an inhibition or enhancement of radiation scattering depending
on the spatial distribution of scatterers, frequency, and parameters
describing each cylinder. As shown in Fig. \ref{fig:2}(c), the optimum structure
barely scatters light and, more importantly, the amplitude of the
scattered field in the near field region is also much smaller than
the amplitude of the incoming plane wave. Hence, we expect to obtain
small multiple scattering effects in an assembly of identical randomly
distributed NW. 

To test this prediction, we numerically calculate scattering efficiencies of NW dimers at different
distances and incident radiation conditions and also the scattered
fields of assemblies of randomly placed NWs. Numerical simulations
were carried out using the RF module of the Finite Element Method
based commercial software COMSOL Multiphysics 4.3. Both in the case
of the scattering properties of the core-shell NW dimers, as in the
simulation of a slab, formed by a random arrangement of these structures,
the geometry of the problem is two-dimensional. In the case of dimers,
the simulation domain consisted on three concentric circles. While
the outer annulus was defined as a Perfectly Matched Layer, in order
to absorb all fields and simulate free space propagation, the first
boundary was used to perform additional calculations. Core-shell NWs
were then defined by two concentric circles placed a distance $d$
one from each other. Material parameters for air and silicon were
taken constant ($\epsilon_{Si}=12.25$ and $\epsilon_{Air}=1.0$ ),
while silver permittivity was taken from  \cite{Johnson_Christy_1972}.
In the case of a slab of randomly arranged NWs, these were distributed
in such a way that covered, at least, the whole range of distances
and orientations studied in the case of dimers. In this case the simulation
domain was a square ($4.2\,\mu\textrm{m}$ side) and scattering boundary
conditions were applied in all exterior boundaries. Scattered field
formulation was used in both cases. This built-in option of the program
allows one to analytically define the excitation field. This was set
to be a plane wave with the electric field pointing out of the simulation
plane. Meshing was done with the program built-in algorithm. The mesh
consisted on triangular elements, with a maximum element size of $42\,\textrm{nm}$
in the case of the random distribution ($20\,\textrm{nm}$ in the
case of dimers), a maximum element growth rate of 1.1, meaning that
adjacent elements to a given one should not be 1.1 times bigger than
it, and a resolution of curvature and narrow regions of 0.2 and 1,
respectively. With these parameters, the meshing algorithm automatically
increases the mesh quality in narrow and curved regions, such as the
core of the wires and the inter-particle spaces. The total number
of elements was 176034 in the, computationally more demanding, case
of randomly distributed NWs. MUMPS direct solver was used, involving
1258331 degrees of freedom, which, in a desktop machine with 4 processors,
required about 3.5 Gb of memory and about 45 seconds per wavelength.
In the case of dimers, scattering efficiency was computed integrating
the outward normal component of the Poynting vector of the scattered
fields in the auxiliary circumference described above, and normalizing
by the incident intensity and geometrical cross section. Results were
tested to reproduce Mie exact solution in the case of an isolated
nanowire.

\begin{figure}[htbp]
\centering\includegraphics[width=1\textwidth]{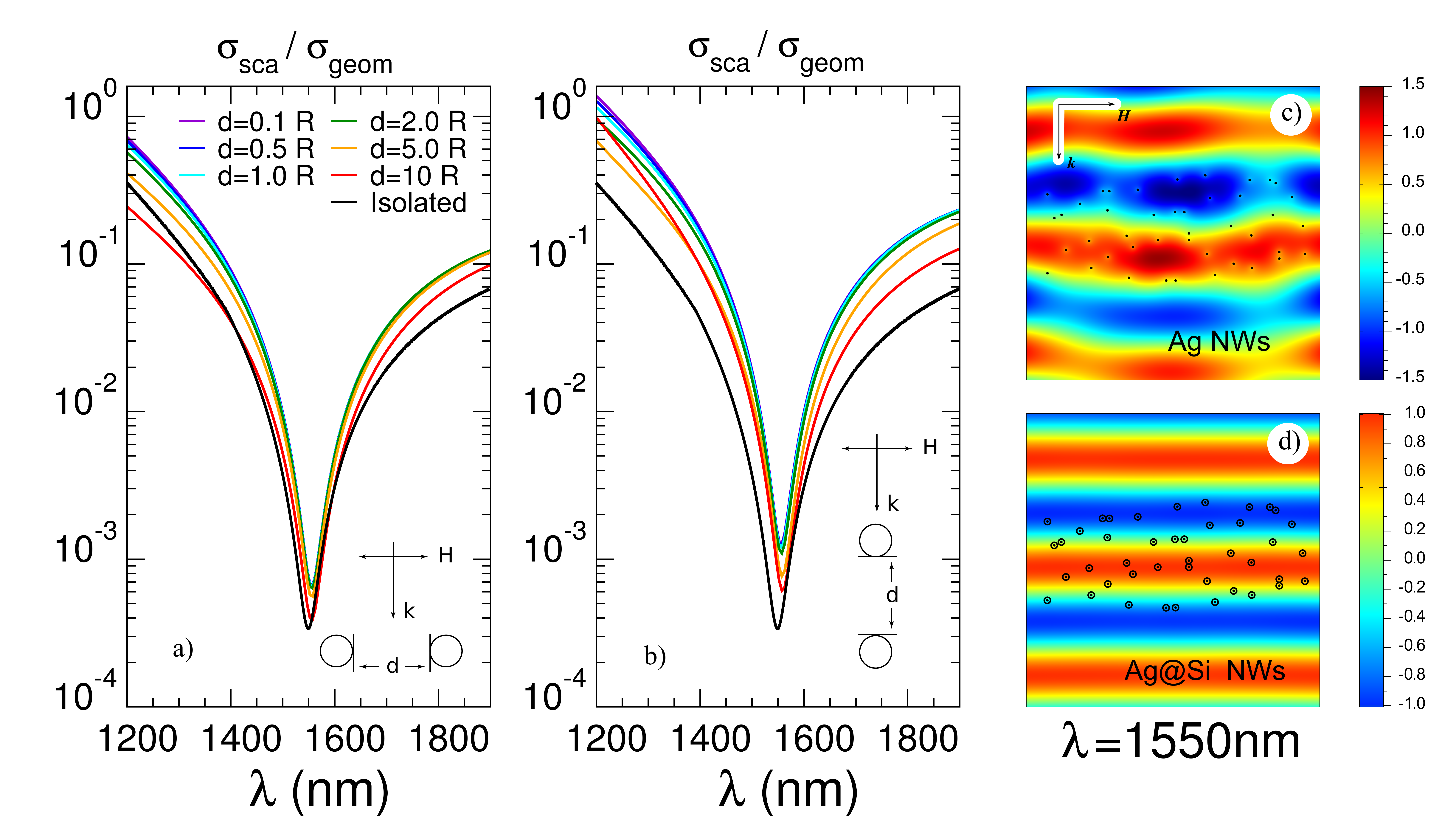}
\caption{\label{fig:4}(a) Scattering efficiency of Ag@Si core-shell
NW dimers ($R_{s}=13.6\textrm{ nm}$, $R_{c}=45\textrm{ nm}$) for
TM polarized plane waves with a wave vector perpendicular to the plane
defined by the dimer for different distances d between the nanowires
(see legends). (b) Scattering efficiency for the same dimers as in
(a), with a wave vector perpendicular to the cylinder axis and in
the same plane as the one defined by the dimer. (c) Map of the electric
field along the cylinder axis direction at a wavelength of $\lambda=1550\textrm{ nm}$
for TM polarized waves for an ensemble of bare Ag NWs ($R=13.6\textrm{ nm}$)
distributed randomly within a slab. (d) Electric field map corresponding
to the the same arrangement of (c). The scattering units in this case
are Ag@Si core-shell NWs ($R_{c}=13.6\textrm{ nm}$, $R_{s}=45nm$)
.}
\end{figure}

 We start by calculating the $Q_{scat}$ of
core-shell dimers. In Fig.  \ref{fig:4}(a) we consider dimers
of equal cylinders with the same parameters as in previous Figs.
For an incoming TM polarization, and normal incidence, $Q_{scat}$
spectra are plotted for different surface-to-surface distance between
cylinders. The range of distances ($d$) varies from $d=0.1\, R_{s}$
to $d=10\, R_{s}$ ($d=4.5\textrm{ nm - }450\textrm{ nm}$). As can
be seen in Fig. \ref{fig:4}(a), the scattering efficiency increases for
distances comparable to the cylinder diameter; nevertheless, $Q_{scat}$
varies only by a factor of two even in the extreme condition of distances
much smaller than the cylinder radius. 

If the wave vector of the incoming plane wave is still perpendicular
to the cylinder axis but contained in the plane defined by the dimer,
the variation of the $Q_{scat}$ with distance between cylinders is
also relatively small: As shown in Fig. \ref{fig:4}(b), $Q_{sca}$ increases
at much by a factor of two when the cylinders are in close proximity.
In this case, a small shift in the position of the minimum of the
scattering spectrum is also observed as a function of the distance.

In order to address a more general case, we have compared the scattering
by an ensemble of Ag@Si to that of bare Ag cylinders. In Fig. \ref{fig:4}(c)
we show a color map of the electric field component parallel to the
cylinders axes for a random ensemble of bare Ag NWs. The incoming
plane wave, at $\lambda=1550\textrm{ nm}$, is TM polarized and its
propagation direction is perpendicular to the cylinder axes (from
top to bottom in the figure). The ensemble  of NWs scatters light rather
strongly: As can be seen in the figure, the transmitted wave is fully
distorted respect to a plane wave. If, for the same positions of the
NW axes and identical incoming plane wave, we consider Ag@Si NWs,
the scattering is inhibited to a large extent as shown in Fig. \ref{fig:4}(d)
where, as previously used, the Ag NW is coated with a $31.4$ nm thick
Si layer. The positions of the NWs were generated randomly within a slab of roughly
$4\mu\textrm{m}$ by $1.5\mu\textrm{m}$ with the sole restriction of keeping a minimum
surface-to-surface distance $d_{min}\simeq 0.1 R_s$.

\begin{figure}[htbp]
\centering\includegraphics[width=1\textwidth]{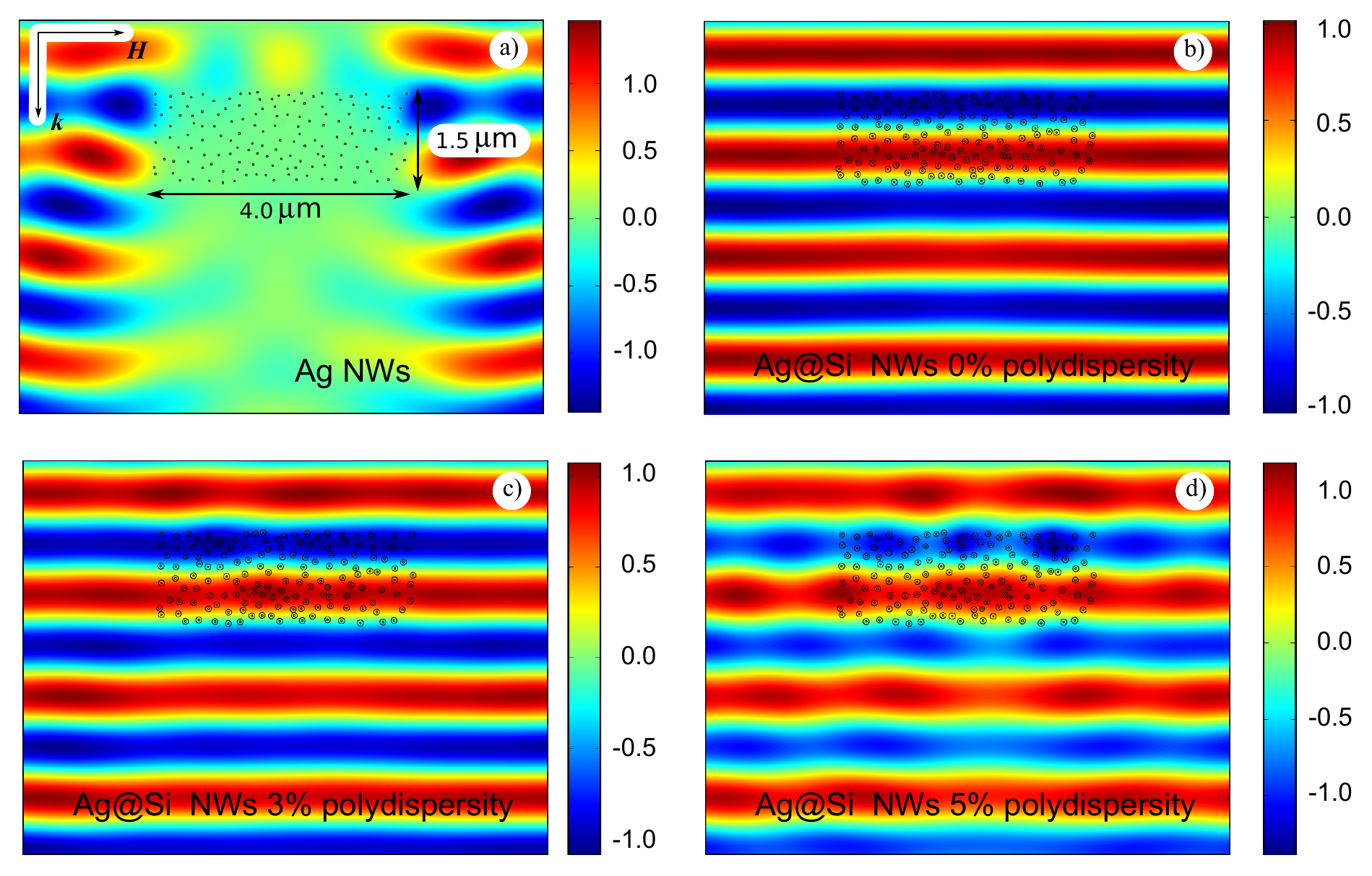}
\caption{\label{fig:5}(a) Map of the electric
field along the cylinder axis direction at a wavelength of $\lambda=1550\textrm{ nm}$
for TM polarized waves for an ensemble of bare Ag NWs ($R=13.6\textrm{ nm}$)
distributed randomly within a slab of $4\mu\textrm{m}$ by $1.5\mu\textrm{m}$. (b) Electric field map corresponding
to the the same arrangement of (a). The scattering units in this case
are Ag@Si core-shell NWs ($R_{c}=13.6\textrm{ nm}$, $R_{s}=45\textrm{ nm}$), the filling fraction of the arrangement is 16\%. (c) Electric field map corresponding to the same structure as
in (b), with the addition of random disorder in both core and shell radii with a standard deviation of 3\% arround the optimal values.
(d) the same as in (c) with a 5\% standard deviation.}
\end{figure}

 Going one step further, we have increased the density of NW in the arrangement and disordered the structural parameters of 
each of the cylinders in the structure. In Fig. \ref{fig:5} we show color maps corresponding to the electric field scattered by different ensembles
of randomly placed NWs occupying a region of dimensions $4\mu\textrm{m}$ by $1.5\mu\textrm{m}$. In Fig. \ref{fig:5}(a), 
the electric field distribution for bare silver ($R=13.6\textrm{ nm}$) NWs is shown. As can be seen, the incoming wave is strongly scattered 
at the working wavelength ($\lambda=1550\textrm{ nm}$). 

In Fig. \ref{fig:5}(b), the same arrangement, in this case of  Ag@Si NWs optimized for transparency, is considered. The filling 
fraction of the NW ensemble is 16\% . Despite this relatively high filling fraction, the electric field map shows that the scattering is 
very small, definitively much smaller than in the case of bare silver NWs.

 In this case, we chose a maximum element size of 4~nm for the silver parts of the system, 10~nm for those of silicon, 
and 80~nm for the whole air domain. In all cases, we chose the same maximum element growth rate (1.1), resolution of 
curvature (0.2) and narrow regions (1.0). These settings generate a mesh with 481574 elements in the case in which no 
polydispersity is assumed. Solution of the system involved 3371719 degrees of freedom, requiring about 6.5 Gb of memory 
and two minutes per wavelength.

 In any realistic application, some fabrication errors are expected. With state-of-the-art techniques, random variations 
of the core and/or shell radii down to a few percents are achievable. In order to consider this kind of fabrication deviations, we have
calculated the light scattering for an assembly of NWs which centers are positioned at the same coordinates 
as in the cases shown in Fig. \ref{fig:5}(a-b) but with
some degree of structural disorder. In Fig. \ref{fig:5}(c), random variations (normally distributed) of the core and shell radii are introduced 
with standard deviations of 3\% for both parameters. As can be seen in ths figure, the wavefronts are slightly disturbed with respect 
to the optimal case. Even if the structural disorder in the NWs is increased up to a 5\% standard deviation, as shown in Fig. \ref{fig:5}(d), 
the transparency condition holds relatively unperturbed.

\section{Conclusions}
We have obtained general conditions under which a core-shell nanowire
presents transparency based on the polarizability with radiative corrections
of the structures. This approach permits us to obtain relatively accurate,
yet simple, analytical expressions for the scattering efficiency spectra
of the systems under study, thereby revealing the interplay among
materials' optical properties and geometry of the NW. These simple
transparency conditions are universal and can be thus extrapolated
to any spectral regime provided that the core-shell cylinders are
relatively subwavelength. 

We have shown that transparency can be achieved with metal@dielectric, 
dielectric@metal and dielectric@dielectric structures
with the appropriate geometrical and optical parameters. Regarding
metal-dielectric cylinders, both metal@dielectric and dielectric@metal structures can be designed
to be transparent under certain conditions. Nevertheless, we have
shown that metal@dielectric structures are much more robust against variations
in the geometrical parameters defining the transparent system. By
contrast, the appearance of localized surface plasmon resonances for
dielectric@metal cylinders close to the transparency condition lead to a large
scattering enhancement for slight NW radius variations (namely, imperfections).

By using an exact Mie-based modeling of the core-shell cylinders,
we have shown that broad-band transparency of metal@dielectric NWs in the infrared
telecommunications range is achievable with realistic materials (i.e.
Si-coated Ag nanowires), even higher than glass. Also, this effect
is shown to be very robust against relatively large geometrical variations
(such as unavoidable fabrication imperfections), and also for a broad
range of angles of incidence and polarization.

Moreover, at the transparency condition, scattering is very weak even
deep in the near field zone. Indeed, our calculations reveal that
an assembly of Ag@Si NWs (even if close to each other) remain highly
transparent.

 Bear in mind that other metals can also be used: i.e. copper, which 
exhibits  an optical response very similar to that of Ag in this spectral 
regime. A wealth of semiconductors with high refractive index and low losses 
could be exploited as well as shells, properly tuning the coating thickness. 

 This system is hence a suitable building block for electrical wiring
where keeping optical transparency is mandatory. Moreover, we anticipate
widespread applications of our theoretical analysis on transparency
in a variety of hybrid metal-semiconductor nanowires, relevant i.e.
to nanophotonics and plasmonics, photovoltaics and metamaterials.

\section*{Acknowledgments}
The authors acknowledge the Spain Ministerio
de Economía y Competitividad, through the Consolider-Ingenio project
EMET (CSD2008-00066) and NanoLight (CSD2007-00046), and NANOPLAS+
(FIS2012-31070) and FIS2012-3611, and the Comunidad de Madrid (grant
MICROSERES P2009/TIC-1476) for support. R. P.-D. and L.S F-P, also
acknowledge support from the European Social Fund and CSIC through
a JAE-Pre and JAE-Doc grants, respectively.

\end{document}